\documentclass{egpubl}

\usepackage[T1]{fontenc}
\usepackage{dfadobe}  

\usepackage{cite}  %
\BibtexOrBiblatex

\electronicVersion
\PrintedOrElectronic

\ifpdf \usepackage[pdftex]{graphicx} \pdfcompresslevel=9
\else \usepackage[dvips]{graphicx} \fi

\usepackage{egweblnk} 

\usepackage[T1]{fontenc}
\usepackage{dfadobe}  
\usepackage{graphicx} %
\usepackage{algorithm}
\usepackage[noend]{algpseudocode}
\usepackage{amsmath}
\usepackage{wrapfig}

\algrenewcommand\algorithmicforall{\textbf{foreach}}
\algrenewcommand\algorithmicindent{.8em}
\usepackage[inkscapeformat=png]{svg}
\usepackage[english]{babel}
\usepackage{makecell}
\usepackage{booktabs} 

\addto\extrasenglish{
  
}
\addto\extrasenglish{
  
}
\addto\extrasenglish{
  
}
\addto\extrasenglish{
  
}
\addto\extrasenglish{
  
}

\usepackage{xcolor}
\usepackage{ulem}%
\usepackage{wrapfig}%

\definecolor{tableheader}{HTML}{CAB476}
\definecolor{tablesubheader}{rgb}{0.79,    0.698,    0.839}
\definecolor{thirdtablecolor}{rgb}{0.8706,    0.7961,    0.8941}
\usepackage{color,colortbl}

\newcommand{\projection}{\ensuremath{\text{Proj}}}

\renewcommand{\emph}[1]{\textit{#1}}

\title{Mesh Simplification For Unfolding}

\author[M. Bhargava C. Schreck M. Freire P.A. Hugron S. Lefebvre S. Sellán B. Bickel]
{\parbox{\textwidth}{\centering 
        M. Bhargava$^{1}$\orcid{0009-0007-6138-6890}
        C. Schreck $^{2}$\orcid{0000-0002-3787-249X} 
        M. Freire $^{2}$\orcid{0000-0002-0926-055X} 
        P.\,A. Hugron $^{2}$\orcid{0009-0001-5592-0723} 
        S. Lefebvre  $^{2}$\orcid{0000-0002-9182-3146} 
        S. Sellán $^{* 3}$\orcid{0000-0003-0003-6807} 
        B. Bickel $^{* 1,4}$\orcid{0000-0001-6511-9385} 
        }
        \\
{\parbox{\textwidth}{\centering 
        $^1$ISTA (Institute of Science and Technology Austria), Austria \\
        $^2$Université de Lorraine, CNRS, Inria, France \\
        $^3$University of Toronto, Canada \\
        $^4$ETH Zürich, Switzerland \\
        $^*$Joint Last Authors
       }
}
}

\begin{document}

\teaser{
    \centering
    \includegraphics[width=\textwidth]{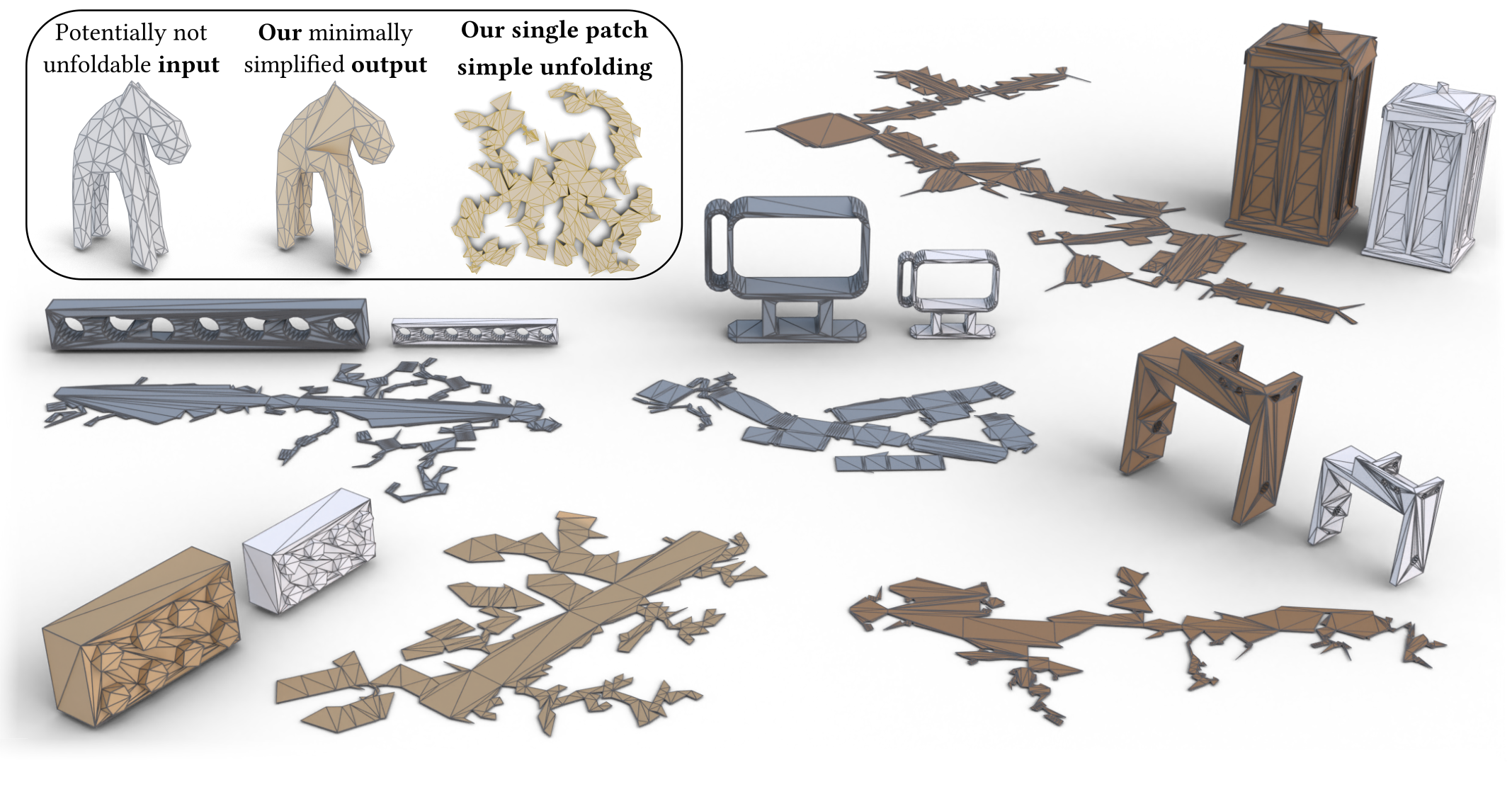}
    \vspace{-1.2cm}
    \caption{Computing a single-patch unfolding for an arbitrary input (corner window, left) can be hard or outright impossible. Instead, we propose relaxing the problem and minimally modifying the input geometry (corner window, center) to make it easily unfoldable (corner window, right). Our algorithm is robust and returns single-patch unfoldings for shapes of arbitrary genus and triangulation quality (see collage). \textbf{See accompanying video for more results.}}
    \label{fig:teaser}
}

\maketitle

\begin{abstract}

We present a computational approach for unfolding 3D shapes isometrically into the plane as a single patch without overlapping triangles.
This is a hard, sometimes impossible, problem, which existing methods are forced to soften by allowing for map distortions or multiple patches.
Instead, we propose a geometric relaxation of the problem: we modify the input shape until it admits an overlap-free unfolding.
We achieve this by locally displacing vertices and collapsing edges, guided by the unfolding process.
We validate our algorithm quantitatively and qualitatively on a large dataset of complex shapes and show its proficiency by fabricating real shapes from paper.

\end{abstract}

\vspace{-0.2cm}
\section{Introduction}\label{section:intro}

An \emph{unfolding} is a map from a surface to a single connected region of the two-dimensional plane that is isometric except for a finite set of discontinuities or \emph{cuts}. If it is bijective (and therefore invertible), it is also called \emph{simple} or \emph{overlap-free} (see examples in \autoref{fig:teaser}). 
Computing simple unfoldings of arbitrary polygonal meshes is a well-known research problem in the geometric fabrication community, as its inverse map provides the necessary instructions to fabricate the shape efficiently by bending pieces of non-stretchable materials like paper, cardboard, wood or Printed Circuit Boards (PCB).

Despite the appeal of this application, finding an overlap-free unfolding of a given input is computationally challenging. Indeed, it is known that many shapes do not even admit such an unfolding \cite{demaine2020acutely}, and even finding if it is possible at all for an arbitrary mesh remains an open problem.

Because of this, existing approaches opt for \emph{relaxing} the problem: for example, by allowing for a limited amount of distortion (i.e., removing the isometric constraint) or by permitting the unfolding to result in several, disconnected patches.
Unfortunately, these relaxations severely hamper the methods' applicability in manufacturing, as many materials cannot stretch without tearing or losing their physical properties (e.g., paper, wood or PCB). Furthermore, assembling an object from a number of unfolded pieces is practically challenging and not even possible for certain fabrication pipelines where single patch unfoldings is a strict requirement like \cite{niu2023pullupstructs} which uses single patch unfoldings to fabricate 3D shapes with pull-up nets. Similarly, \cite{pcbend2023} required their unfoldings to be single patch to maintain electrical connectivity in their 3D circuits. 

\begin{figure*}
    \centering
    \includegraphics[width=\linewidth]{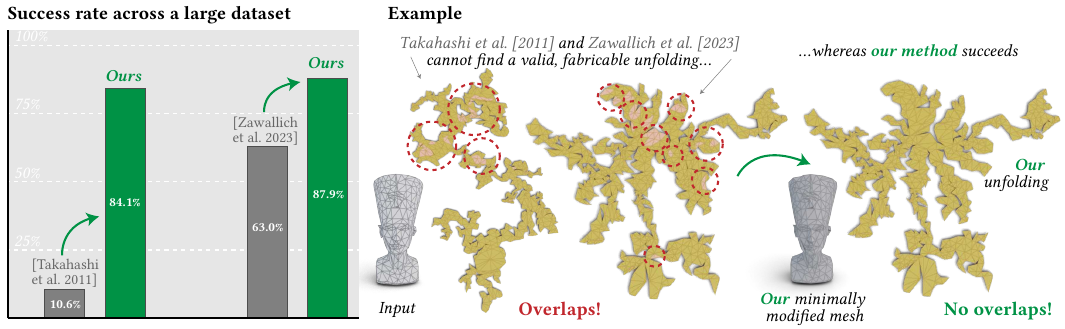}
    \vspace{-0.75cm}
    \caption{Existing methods fail at consistently producing simple, non-overlapping unfoldings of complex shapes (middle). Instead, our method finds a close approximation to the input shape that admits a simple unfolding (right). In practice, this massively increases the number of shapes that can be fabricated from a single sheet of material (see quantitative bar chart on the left).}
    \label{fig:related-fig}
     \vspace{-0.3cm}
\end{figure*}

Instead, we propose a \emph{geometric} relaxation of the problem, modifying the input until it can be unfolded in a single patch.
Ironically, our reframing takes advantage of the same complexity of the space of foldable shapes that has thwarted previous efforts: indeed, given an input mesh without a bijective unfolding, it is likely that there exists a \emph{very similar} shape which \emph{is} easily unfoldable in a single patch.
Through a combination of local vertex displacements and guided edge collapses, our algorithm is a robust mesh processing strategy that returns a simple unfolding without making any assumption about the topological properties or triangulation quality of the input (see \autoref{fig:teaser}).

We evaluate our algorithm on a vast array of qualitatively representative examples for which previous works are unable to find overlap-free unfoldings (see \autoref{fig:related-fig}). We quantitatively confirm these findings through large-scale experiments, which we also use to validate our parameter choices and ablate different algorithmic steps. Our method can be readily applied to common manufacturing frameworks, as we showcase with our fabricated results. Finally, we prototype the incorporation of user-defined constraints and end by discussing potential avenues for future work.

\vspace{-0.1cm}
\section{Related Work: Simple Unfoldings}
\label{section:related}

Finding a simple unfolding of a surface is a core challenge faced by methods fabricating from planar sheets, whether to create origami by folding paper \cite{demaine2002recent,callens2018flat,Polthier2009ImagingM}, to laser-cut objects that can be pulled up with strings \cite{niu2023pullupstructs}, bend shape memory composites \cite{felton2013self}, to bend PCBs to manufacture surface electronics \cite{pcbend2023} or to make robots using robogami \cite{schulz2017interactive,yao2019reconfiguration,BelkeThesis2020}. 
Thus, it has been extensively studied.

Beyond basic shapes like spheres \cite{van2008unfolding} and other convex surfaces \shortcite{starUnfolding97}, computing an overlap-free unfolding of a general input surface (if it exists at all \cite{demaine2020acutely}) remains an unsolved open problem. Thus, existing approaches resort to \emph{relaxing} the problem, computing mappings from surfaces to the plane that satisfy only most of the properties of a simple unfolding.

\subsection{Relaxations: Distorted and Multi-Patch Unfoldings}

A common relaxation of the problem is no longer forcing the map to preserve the metric of the surface, instead allowing it to introduce a minimal amount of distortion. This more general class of maps are known as \emph{parametrizations} of the surface, and their computation has long been a commonly considered task in the Computer Graphics community due to its application to texture mapping. While a comprehensive review of the study of surface parametrizations is beyond the scope of this paper (see surveys by \cite{wei2010survey} and \cite{fu2021inversion}), it suffices to say that great progress has been made: e.g., in obtaining conformal~\cite{levy2023least, sawhney2017boundary}, as rigid as possible~\cite{liu2008local} or low distortion parametrizations \cite{Fargion2022globally, su2020efficient, smith2015bijective, jiang2017simplicial} or in efficiently optimizing for both cut placement and distortion~\cite{li2018optcuts, poranne2017autocuts}.

However, such low distortion mappings are not sufficient for manufacturing applications like those listed at the beginning of this section. Materials like wood, paper or metal are truly non-stretchable, thus requiring \emph{strictly isometric} unfoldings. Thus, a more relevant relaxation of the problem is allowing the unfolding to result in more than one patches or disconnected regions.

A popular approach is to make the shape piecewise developable \cite{stein2018developability,sellan2020developability}, where each segment can be mapped isometrically to 2D \cite{lawrence2011developable,yuan2023survey}. \cite{tachi2009origamizing} showed how to fold a piece of paper into a 3D shape using tuck folds, an approach turned by \cite{demaine2017origamizer} into a more practical algorithm. Although this works on many complex meshes, fabricating objects with tuck folds requires a lot of extra paper as well as significant folding time.

To simplify manufacturing, an alternative strategy assumes that the input shape is represented via a triangle mesh (which is trivially piecewise-developable) and that the cuts in the unfolding are restricted to coincide with mesh edges \cite{demaine2007GeometricFoldingAlgorithms, o2019unfolding}. This class of mappings, commonly known as \emph{edge unfoldings}, is the one we concern ourselves with.

\begin{figure}
    \centering
    \vspace{-0.45cm}
    \includegraphics[width=\linewidth]{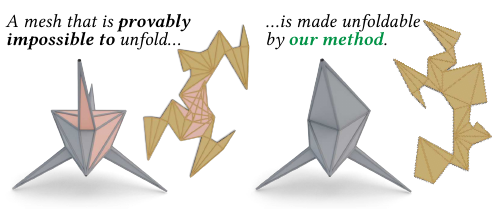}
    \vspace{-1.0cm}
    \caption{Not only are simple unfoldings hard to compute; in some cases, they can be theoretically shown not to exist \cite{demaine2020acutely}. This motivates us to find a close approximation to the input shape that admits an unfolding.}
    \label{fig:demaine}
    \vspace{-0.35cm}
\end{figure}

\subsection{Edge Unfoldings of Triangle Meshes}

\begin{figure*}
    \centering
    \includegraphics[width=\linewidth]{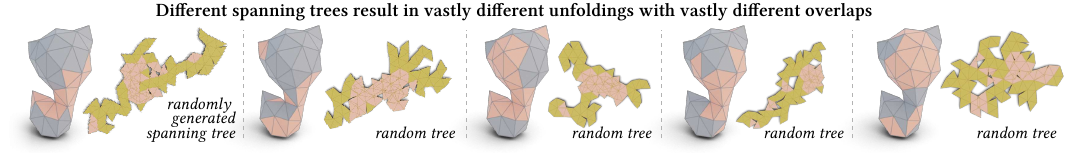}
    \vspace{-0.75cm}
    \caption{The potential single-patch unfoldings of a given mesh can be represented by the spanning trees of its dual graph. The choice of spanning tree (here, random) can greatly affect the shape of the unfolding and the number of overlaps in it.}
    \label{fig:spanning-trees}
     \vspace{-0.3cm}
\end{figure*}

For a general shape represented as a 3D mesh, finding a single-patch edge unfolding boils down to generating the dual graph and finding a spanning tree over it.
When a mesh has a few hundred triangles a random spanning tree usually results in a single-patch unfolding with multiple overlaps~\cite{Schevon1989} (see \autoref{fig:spanning-trees}).

Finding an edge unfolding consisting of a minimal number of disconnected patches is computationally hard \cite{demaine2007GeometricFoldingAlgorithms}. Thus to obtain a single-patch edge unfolding many methods rely on heuristics and split the final unfolding into several patches to get rid of any remaining overlaps \cite{schlickenrieder1997nets}.

Agarwal et al.~\shortcite{starUnfolding97} suggest star unfoldings to obtain a single-patch unfolding, which works for general unfoldings of convex shapes but fails to find a solution for edge unfoldings. 
The minimum perimeter heuristic \cite{straub2011creating} minimizes the number of cuts used to obtain an initial unfolding and then resolves all overlaps by adding further cuts to separate them into multiple patches by solving a minimum set cover problem. 
Takahashi et al.~\shortcite{takahashi2011optimized} introduce a topological surgery approach where the input mesh is divided into smaller patches that are later merged using a genetic algorithm to obtain an unfolding. If it fails to generate a single-patch unfolding, the method outputs multiple overlap-free patches.
Korpitsch at al.~\shortcite{korpitsch2020simulated} use simulated annealing on the different spanning trees to obtain an unfolding with fewer overlaps. 

Addressing a similar goal, Zawallich~\shortcite{Zawallich2023} demonstrated empirically that Tabu search can give a significant improvement in finding an unfolding with real-time performance for moderate mesh complexities. 
In a concurrent work, Zawallich~\shortcite{zawallich2024unfolding} used their unfolder in combination with different existing surface flows methods to showcase with great success unfolding of genus-zero meshes.
Unfortunately, none of these algorithms can generate simple unfoldings on general input shapes with arbitrary genus, nor do they suggest a method to approximate the mesh to obtain one (see \autoref{fig:related-fig}).

\begin{figure}
    \includegraphics[width=\linewidth]{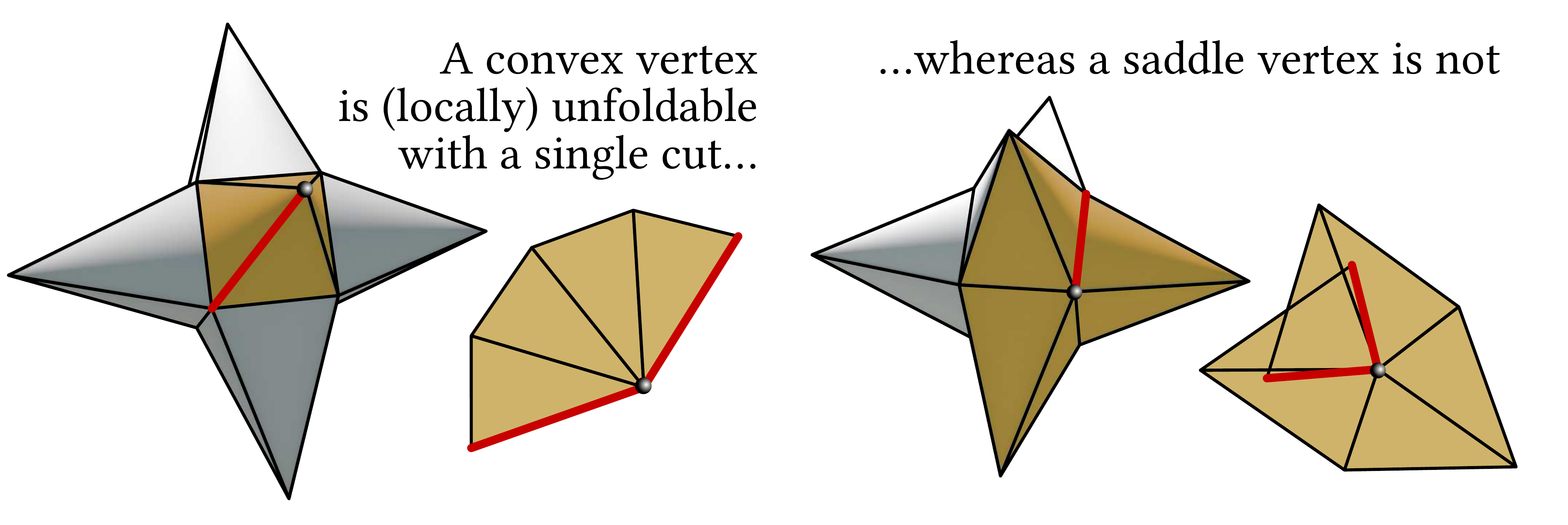}
    \vspace{-0.35in}
    \caption{A convex vertex unfolds in 2D without overlaps with one cut, but a saddle vertex's single cut always causes overlapping triangles.}
    \label{fig:vertex-types}
\end{figure}

Interestingly, Demaine et al.~\shortcite{demaine2020acutely} noted that an unfolding being overlap-free is not a topological property but is connected to the geometry of the shape (see \autoref{fig:demaine}). This matches our observation that even small changes to the geometry can greatly reduce overlaps, and turn a challenging case for which no solution is found into a feasible one. We propose taking advantage of this through a \emph{geometric} relaxation of the problem: starting from a potentially overlapping initial isometry, we progressively modify the input geometry and this map to achieve a single-patch overlap-free unfolding.

\vspace{-0.2cm}
\section{Motivation: \textit{Not-so-simple} unfoldings}
\label{section:motivation}

Let us briefly consider the problem of finding an exact, single-patch overlap-free (or \textit{simple}) unfolding $\Gamma$ of a given input mesh $\Omega$ with vertices $\mathcal{V} = \{v_1,\dots, v_n\}$, triangular faces $\mathcal{T} = \{t_1,\dots,t_m\}$ and (undirected) edges $\mathcal{E} = \{e_1,\dots,e_k\}$. 
The space of all single-patch unfoldings of $\Omega$ is explored through the mesh's \textit{dual graph}, i.e.\ the graph whose nodes correspond to mesh faces and in which any two faces are connected with a link if they share a mesh edge. For consistency, we will use \textit{edge} and \textit{vertex} to refer to the triangle mesh's elements and \textit{node} / \textit{link} for the dual graph's elements, to avoid confusion.

\begin{wrapfigure}[10]{r}{1.4in}
      \vspace{-0.2in}
      \hspace{-0.25in}
         \includegraphics[width=1.6in]{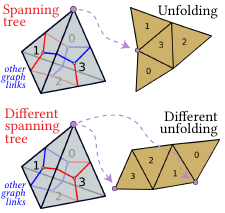}
\end{wrapfigure}
Critically, each \textit{spanning tree} (i.e.\ a connected sub-graph that includes all nodes without links creating a cycle) of the dual graph corresponds to a possible unfolding of the mesh: one need to only choose an arbitrary start face (node) and rigidly transform it onto the plane, then use the shared mesh edges as \textit{hinges} to unwrap the faces that are linked to it in the tree (see inset) until all faces are exhausted. Every mesh edge that is not used as a hinge is commonly referred to as a \textit{cut} edge, since these are the edges that should be physically cut to fabricate the object.

\begin{figure*}
  \vspace{-0.2cm}
  \includegraphics{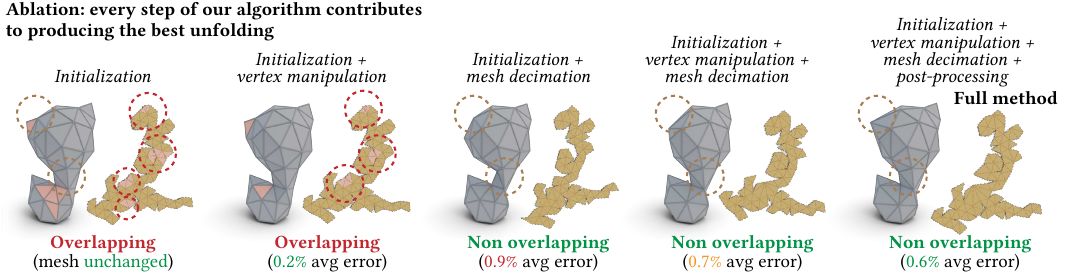}
  \vspace{-0.75cm}
  \caption{Our algorithm is composed of several steps, all of which are crucial to obtaining a single patch unfolding that is non-overlapping (see first two subfigures) and differs as little as possible from the original mesh (see regions circled in brown).}
  \label{fig:ablation}
  \vspace{-0.3cm}
\end{figure*}

Up to an arbitrary rigid global transformation, this process ends with a rigid mapping associated to every mesh face $R_1,\dots,R_m$ that unfolds it onto the plane. By extension, this also creates a correspondence between each mesh vertex $v_i$ and its potentially multiple 2D images $u_{i,1},\dots,u_{i,n_i}$. (see the inset above, where the purple vertex has one copy in 2D in the first unfolding but two in the second).

Unfortunately, for a general input mesh and a general spanning tree, the unfolding will contain a large number of overlaps (i.e.\ intersections among the unwrapped faces, highlighted in \autoref{fig:spanning-trees}).

Sometimes, these overlaps can be explained \textit{locally} by simple geometric reasons: for example, as explored by \cite{takahashi2011optimized}, any vertex neighborhood with non-zero angle deficit cannot be unfolded onto the plane without cuts (as shown in \autoref{fig:vertex-types}, locally convex or concave vertices will require at least one cut, whereas saddle-like ones will require at least two). 
\begin{wrapfigure}[8]{r}{0.65in}
      \vspace{-0.10in}
      \hspace{-0.20in}
         \includegraphics[width=0.75in]{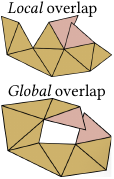}
\end{wrapfigure}
Apart from local overlaps, an unfolding may also have \emph{global} overlaps (see inset), which are significantly more challenging to identify as they require explicit checking for triangle overlaps.

Exploring the entire space of spanning trees (and thus of mesh unfoldings) to find an overlap-free one explodes in combinatorial complexity and rapidly becomes computationally intractable. Interestingly (for a mathematician) but disappointingly (for a manufacturer), an overlap-free unfolding may not exist even for relatively simple geometries (see \autoref{fig:demaine}).
Instead of following the approach of previous works and softening the problem statement by allowing minimal overlaps or more than a single disconnected patch, we propose a paradigm shift and instead consider simplifying the \emph{mesh geometry} to ensure a single-patch unfolding.

However, standard simplification methods (e.g., \cite{QEMDecimation1997}) are not guaranteed to resolve or even reduce the unfolding's overlaps. Thus, we propose an overlap-aware strategy to simplify a mesh while ensuring that each step reduces the overlap.

\vspace{-0.1cm}
\section{Method: Mesh Simplification for Unfolding} \label{section:main-method}

In lieu of the difficulties stated in \autoref{section:motivation}, we now update our problem statement to that of finding a \emph{new} mesh $\Omega^\star$ similar to the input $\Omega^0$ but having a single-patch non-overlapping unfolding $\Gamma^\star$ . 
We start with an initial, potentially overlapping, unfolding $\Gamma^0$ 
(we will discuss initialization strategies in \ref{section:initialization}). We then iteratively alternate between reducing the overlaps, first \emph{geometrically}, by manipulating the vertex positions of the mesh  (\autoref{section:vertex-repositioning}) and then, \emph{topologically}, through a purpose-built unfolding-aware mesh decimation strategy (\autoref{section:mesh-simplification}). Once we achieve an overlap-free unfolding, we post-process the obtained mesh to ensure it is as close as possible to the input without introducing any new overlaps (\autoref{section:postprocessing}).
The effect of each individual step is showcased in \autoref{fig:ablation}, and the overall approach is summarized in \autoref{algorithm:algorithm-main} in the Supplemental.

\subsection{Unfolding-aware vertex manipulation}
\label{section:vertex-repositioning}

At a given iteration $t$, we call the current mesh $\Omega^t$ and its unfolding $\Gamma^t$. We first seek to remove small overlaps that can be handled by only displacing vertices without changing the topology of the mesh. 
Experimentally, we distinguish between three classes of overlaps in the unfolding $\Gamma^t$ for which overlaps may be resolved geometrically by slightly repositioning its mesh vertices. 

\begin{wrapfigure}[5]{r}{0.6in}
    \label{need to update the text of this figure.}
      \vspace{-0.25in}
      \hspace{-0.24in}
         \includegraphics[width=0.8in]{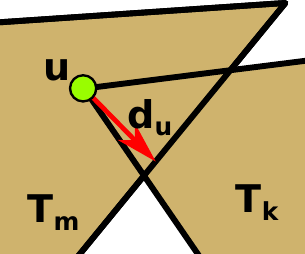}
\end{wrapfigure}

\textbf{Case 1:} In the unfolding $\Gamma^t$, if a vertex $u_{i,j}$ of a triangle $T_k$ falls inside another, topologically non-adjacent triangle $T_m$ (see inset) we find $p$, the closest point to $u_{i,j}$ just outside of the boundary of $T_m$ and compute the vector $d_u$ that would remove the overlap between triangles $T_k$ and $T_m$ $\Gamma^t$. We then lift the vector $d_u$ from the unfolding to the mesh $\Omega^t$ with the help of barycentric coordinates to obtain $d_v$. We displace the three-dimensional vertex $v_i$ with $d_v$ thus moving its unfolding vertex $u_i$ to $p$). While this process will resolve the given overlap, the effect of displacing $v_i$ is not limited to $u_{i,j}$, but results in a displacement in every other planar image of $v_i$, $u_{i,s}$, $s\neq j$ which may result in new overlaps. If this were the case, we skip this operation.

\textbf{Case 2:} If we identify an overlap between two topologically adjacent triangles that share a common non-saddle vertex, we resolve this overlap by moving the corresponding 3D non-saddle vertex in the direction opposite to the vertex normal. This operation flattens the vertex locally in its 1-vertex neighborhood, thus reducing the curvature around that vertex. Since all triangles attached to this vertex contract, no new triangle overlaps are created. We contract the vertex just enough to resolve the overlap via adaptive stepping.

\textbf{Case 3:} Finally, we attempt to reduce the overlaps between all pairs of overlapping triangles that are only overlapping with one another, by following the gradient flow of a geometric energy
\begin{equation}\label{equ:collision}
    \Phi (v_1,\dots,v_n) = w_c\Phi_c+w_s\Phi_s
\end{equation}
where $w_c$ and $w_s$ are weights, $\Phi_c$ is a standard collision energy and $\Phi_s$ is an elastic regularizer. Details on the computation of $\Phi_c$ and $\Phi_s$ can be found in \autoref{section:surface-flow} in the Supplemental.

\subsection{Unfolding-aware mesh decimation}
\label{section:mesh-simplification}

Likely, the vertex manipulations described in \autoref{section:vertex-repositioning} do not remove all existing overlaps, forcing us to resort to more invasive mesh surgery.
Inspired by the classic work of \cite{QEMDecimation1997}, we collapse mesh edges sequentially by drawing them from a priority queue. In our case, our queue is populated by all edges that participate in any overlaps, and their priority value is given by the integer number of other triangles they are overlapping. 

\begin{figure}
\vspace{-0.3cm}
    \includegraphics{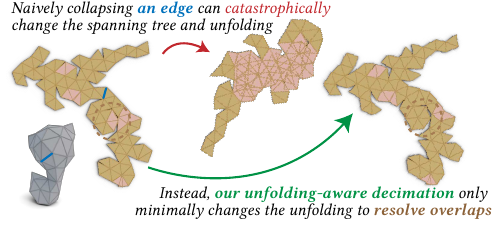}
    \vspace{-0.35cm}
    \caption{Through the strategy described in the text, we repair the spanning tree after each edge collapse. This removes the need to recompute the unfolding after each collapse and ensures that the new unfolding is only minimally different from the pre-collapse one.}
    \label{fig:preserve-spanning-tree}
    \vspace{-0.3cm}
\end{figure}

Given that an edge collapse is a change in the dual graph (specifically, it eliminates two nodes and a maximum of five links, creating two new links), it may cause the spanning tree to cease being a tree (by no longer being connected) or cease being a spanning one (by no longer including every node), thus no longer corresponding to a valid mesh unfolding.
Thus, a naive mesh decimation strategy would require recomputing a spanning tree (i.e.\ a new unfolding) after every collapse. This would be computationally prohibitive and, given the diversity in possible unfoldings, it would make our simplification algorithm highly unstable (see \autoref{fig:preserve-spanning-tree}). 

\begin{figure}
\vspace{-0.1cm}
    \includegraphics[width=0.9\linewidth]{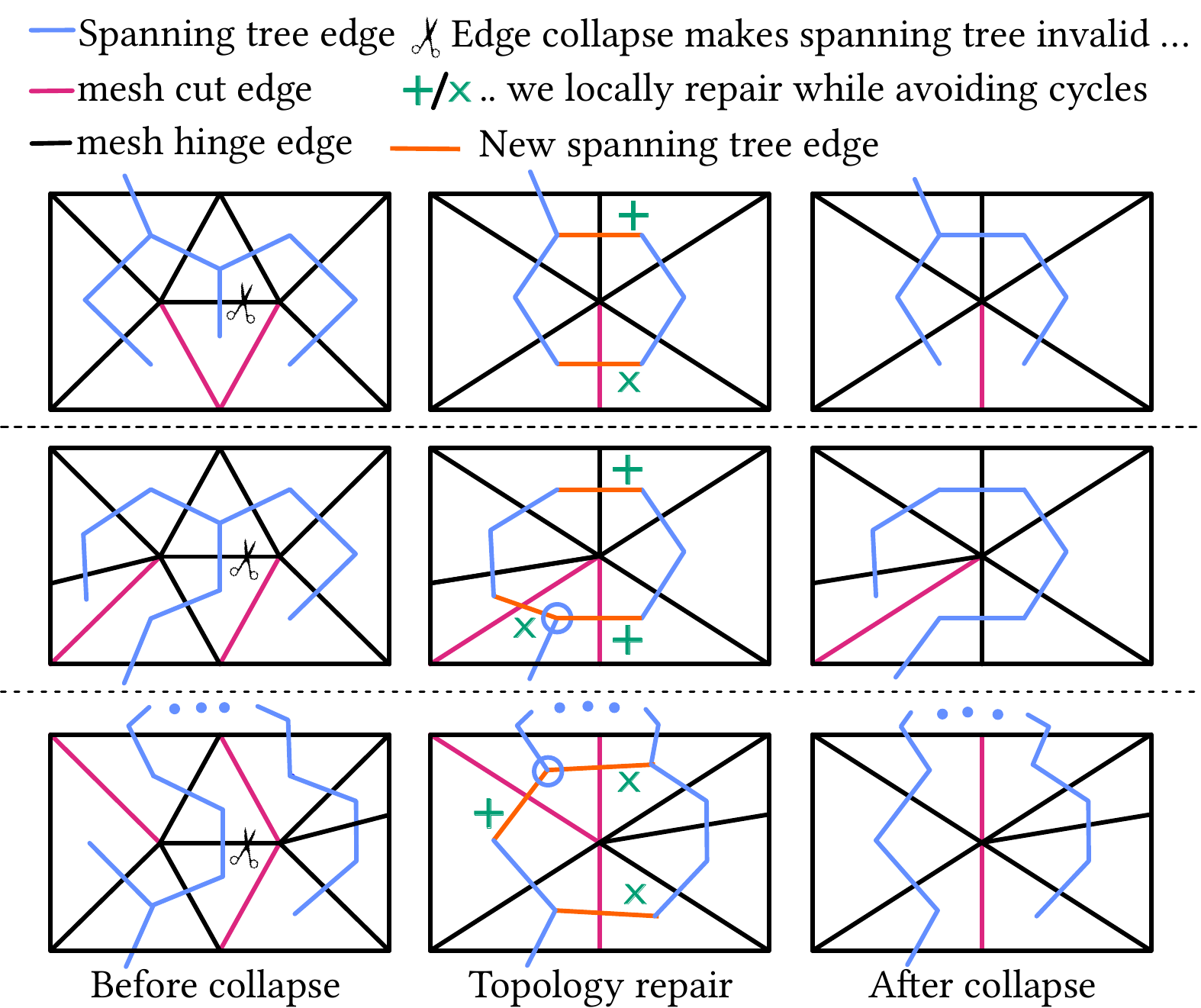}
    \vspace{-0.3cm}
    \centering
    \caption{Illustration of the spanning tree-preserving edge collapse operation on three examples, one per row. 
    }
    \label{fig:spanning-tree-preserve-explain}
    \vspace{-0.3cm}
\end{figure}

To avoid this, we instead propose preserving the spanning tree during the collapse. Fortunately (see  \autoref{fig:spanning-tree-preserve-explain}), this can be done in constant complexity by looping over every node in the edge's one-ring, and adding as many of the links in it to the spanning tree as possible without introducing a cycle. At the end of this loop, we are guaranteed to still be in possession of a single-component spanning tree. 
If the number of overlaps caused by this tree's unfolding is equal or higher than before the collapse, we revert the collapse and proceed to the next element in the queue. As is common in re-meshing algorithms, we similarly check if the collapse caused non-manifold vertices, self-intersections or flipped triangles, in which case we also revert it. 
Notably, this construction is independent to the placement of the vertex resulting from the collapse. In practice, we consider the midpoint as well as at the original edge's vertices, and choose whichever one results in the lowest number of overlaps.

We repeat this decimation process until the queue is exhausted or every edge in the queue is rejected, in which case we begin a new iteration of our algorithm, starting from the unfolding-aware vertex manipulation of \autoref{section:vertex-repositioning}. Pseudocode for our full unfolding-aware decimation strategy is provided in Algorithm~\ref{algorithm:decimation} in the Supplemental.

\textit{Exit strategy:} The iterative approach described above can get \emph{stuck} if every edge in the initial collapse queue is rejected (i.e.\ because its collapse results in degenerate feature, which could cause us to skip over the decimation step altogether. Experimentally, we find our algorithm getting stuck only in later iterations after the vast majority of overlaps have been resolved but a small minority remain. In these cases, we rely on two heuristic exit strategies that ensure progress in the algorithm: \textit{broadening our search} (by reinitializing the priority queue to include the neighbors of overlapping edges) and \textit{lowering our expectations} (by choosing the collapse which causes the lowest number of new overlaps ).

We repeat this outer iteration loop until we obtain a single-patch, non-overlapping unfolding (a \emph{success} by our algorithm) or we exhaust a maximum number of iterations (in practice, we use $100$) without finding one (a \emph{failure}).

\subsection{Post-processing}
\label{section:postprocessing}
If successful, the iterative approach described in Sections \ref{section:vertex-repositioning} and \ref{section:mesh-simplification} terminates with a mesh $\Omega$ and its single-patch, non-overlapping unfolding $\Gamma$. However, in achieving this unfolding, we may have deviated unnecessarily from the input $\Omega^0$. 
As a final step, we search to optimize the position of the vertices of $\Omega$ in order to reduce the distance between $\Omega$ and $\Omega^0$ while avoiding creating new overlaps.

Experimentally, we find that optimizing the vertices of $\Omega$ while strictly avoiding overlaps is unnecessarily restrictive. Instead, we observe that more significant improvements in accuracy can be achieved by allowing for temporary, small overlaps in $\Gamma$ to appear during the optimization and subsequently removing them.
Because of this, we advocate for a post-processing strategy that \emph{decouples} the mesh from its unfolding and instead links the two through a coupling energy term, simultaneously penalizing overlaps and deviation from $\Omega^0$.
We do this by considering $v_i$ and $u_i$ as separate variables, and iteratively optimizing the energy
\begin{equation}
    E_p(\Omega,\Gamma) = \delta_d E_d(\Omega,\Omega^0) + \delta_d E_o(\Gamma) + \delta_c E_c(\Omega,\Gamma)
\end{equation}
where $E_d$ is a distance energy between $\Omega$ and $\Omega^0$, $E_o$ measures the overlap in $\Gamma$, and $E_c$ couples the two (see \autoref{section:post-processing-appendix} in Supplemental for details on the construction of these energies). We optimize using a gradient descent. $\delta_d$, $\delta_o$, and $\delta_c$ are boolean terms that allow for scheduling the energies: we first optimize only $E_d$ to bring $\Omega$ closer to $\Omega^0$, and then alternate between optimizing $E_o$ to remove overlaps in $\Gamma$, and $E_c$ to ensure that the output remains a viable unfolding. If this is not achieved, we make the method output the last valid single-patch, non-overlapping unfolding (e.g., the output of \autoref{section:mesh-simplification}).

  \begin{table*}
   \vspace{-0.3cm}
    \begin{center}
    \caption{
        When ran on large-scale datasets generated from Thingi10K \cite{zhou2016thingi10k}, our algorithm significantly increases the success rate of the methods by \cite{takahashi2011optimized} and \cite{Zawallich2023} at finding a simple, single-patch unfolding.}
    \vspace{-0.3cm}
    \label{table:many-shapes-experiment}
    \resizebox{\linewidth}{!} 
    {
    \addtolength{\tabcolsep}{-0pt}
    \renewcommand{\arraystretch}{1.1} %
    \begin{tabular}{l||c|c|c|c||c|c|c|c}
\toprule
\rowcolor{tableheader}
Dataset & Success (Takahashi) & Runtime (Ta.) & Success (Ta. + \textbf{Ours}) & Runtime (\textbf{Ours}) & Success (Zawallich) & Runtime (Za.) & Success (Za. + \textbf{Ours}) & Runtime (\textbf{Ours}) \\
\midrule
\emph{coarse} &  40.24\% & 24.5 s & \textbf{92.42\%} & 2.77 s & 85.30\% & 9.8 s & \textbf{91.96\%} & 11.69 s \\
\rowcolor[HTML]{EFEFEF} \emph{fine} & 10.55\% & 111.7 s & \textbf{84.10\%} & 18.64 s & 62.99\% & 30.7 s & \textbf{87.93\%} & 14.52 s \\
\midrule
\end{tabular}

    }
    \end{center}
  \end{table*}
\begin{figure*}
    \vspace{-0.35cm}
    \centering
    \includegraphics[width=7.1in]{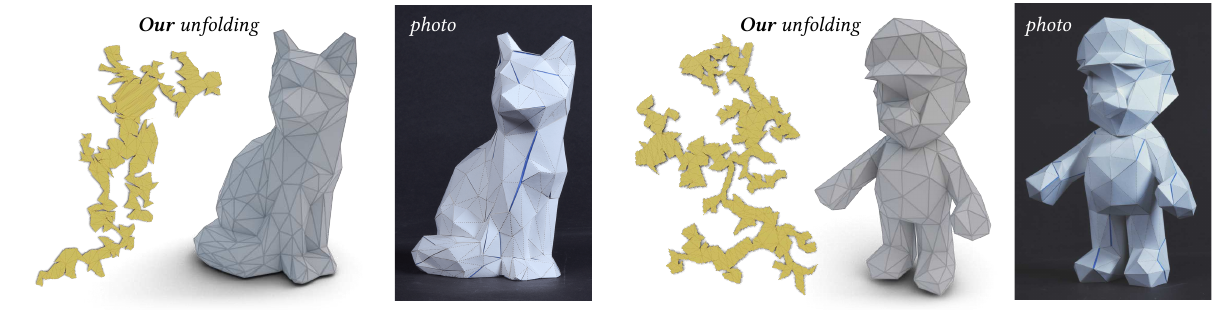}
    \vspace{-0.85cm}
    \caption{We carry out the clearest validation of our algorithm by utilizing it to generate physical models of a fox and an Italian plumber from pieces of cut paper. Both shapes were originally non-unfoldable by \cite{takahashi2011optimized}.}
    \vspace{-0.35cm}
    \label{fig:fabrication-results}
    \end{figure*}

\subsection{Implementation details}\label{section:initialization}

We implemented our method in C$++$, using Libigl \cite{libigl} and CGAL \cite{fabri2009cgal} for common geometric processing subroutines. We report runtimes carried out on a Linux machine with 64GB memory and a i9-9900K \@ 3.60GHz processor.

\textit{Initialization. }
Like any other optimization method, our approach is also sensitive to the initial (overlapping) unfolding:
broadly, the fewer overlaps present in our initialization, the more likely our algorithm is to successfully terminate within a small distance of the input mesh.
Thus, instead of opting for a random or heuristic based initial unfolding, we use two different search based methods: the unfolder by proposed by \cite{takahashi2011optimized} and the one proposed by \cite{Zawallich2023}.

On complex examples, Takahashi et. al.'s genetic algorithm may fail to find a single-patch unfolding, in which case it generates multiple patches of overlap-free unfolding. We merge these separate patch into a single (overlapping) patch by sorting them by size in decreasing order, experimenting with all the possible edges through which each patch can be merged and selecting the one that causes the lowest number of overlaps. On the other hand, Zawallich et al.'s Tabu-search based method uses an efficient data structure to find the least overlapping unfolding, making its output readily available for our method.

Our method remains agnostic to any specific initialization method and can seamlessly incorporate new initialization methods should future algorithms present improvements.

\begin{table*}
    \begin{center}
    \vspace{-0.4cm}
    \caption{Quality metrics of our method on a gallery of shapes from the Stanford 3D Scanning Repository.}
    \vspace{-0.3cm}
    \label{table:statistics-gallery}
    \resizebox{1.0\linewidth}{!} 
    {
    \addtolength{\tabcolsep}{7.5pt}
    \renewcommand{\arraystretch}{1.2} %
    \begin{tabular}{l||c|c|c|c|c|c}
    \toprule
    \rowcolor{tableheader}
    Mesh name & Input face count & Initialization  & Output face count  & Hausdorff distance   & Chamfer distance & Runtime \\
    \midrule
    Armadillo  & 1200 & \cite{takahashi2011optimized} & 1164 &  0.040 & 0.0013 & 2.38 s\\
    \rowcolor[HTML]{EFEFEF}Armadillo  & 1200 & \cite{Zawallich2023} & 1192 &  0.023 & 0.0004 & 2.53 s \\
    \midrule
    Bunny  & 1000 & \cite{takahashi2011optimized} & 956 &  0.0098 & 0.00103 & 6.78 s \\
    \rowcolor[HTML]{EFEFEF}Bunny  & 1000 & \cite{Zawallich2023} & 998 &  0.0016 & 0.00008 & 0.66 s \\
    \midrule
    Bishop  & 496 & \cite{takahashi2011optimized} & 488 &  0.011 & 0.00078 & 0.64 s\\
    \rowcolor[HTML]{EFEFEF}Bishop  & 496 & \cite{Zawallich2023} & 490 &  0.011 & 0.00073 & 1.32 s \\
    \midrule
    Nefertiti  & 1000 & \cite{takahashi2011optimized} & 930 &  0.017 & 0.0011 & 11.85 s \\
    \rowcolor[HTML]{EFEFEF}Nefertiti  & 1000 & \cite{Zawallich2023} & 950 &  0.011 & 0.0008 & 10.20 s\\
    \midrule 
    Plane  & 1200 & \cite{takahashi2011optimized} & 986 &  0.017 & 0.0016 & 27.53 s \\
    \rowcolor[HTML]{EFEFEF}Plane  & 1200 & \cite{Zawallich2023} & 1070 &  0.008 & 0.0011 & 55.57 s\\
    \midrule
    \end{tabular}
    }
    \end{center}
    \vspace{-0.4cm}
  \end{table*}
  
\vspace{-0.2cm}
\section{Results}\label{section:results}

Our algorithm is motivated by a very practical problem: that of unfolding a given 3D shape so that it can be fabricated from a planar material.
Thus, beyond the theoretical observations made in \autoref{section:main-method}, we first and foremost evaluate our algorithm practically, by justifying every algorithmic and parametric choice as well as its superior performance over previous works on a vast array of inputs.

In order to quantitatively evaluate these aspects of our method and others, we produce two datasets of meshes obtained by decimating the Thingi10K dataset \cite{zhou2016thingi10k} to 500 and 1000 faces, respectively.
After pruning the dataset to remove non-manifold, self-intersecting, and multi-component meshes, we are left with \textit{3049} meshes with 500 faces (which we refer to as our \emph{coarse} dataset) and \textit{2440} meshes with 1000 faces (our \emph{fine} dataset).

The main metric we use to evaluate our method and others if it is \emph{successful} or not is defined as whether the algorithm finds a single-patch, non-overlapping unfolding. On every success of our method, we also measure and report the time taken to find a solution, the number of faces in the output, as well as the quality of the final mesh approximation, given by the Hausdorff and Chamfer distances between the input and output meshes. 

\vspace{-0.1cm}
\subsection{Experiments}

\begin{figure*}
\centering
\vspace{-0.2cm}
\includegraphics[width=0.92\linewidth]{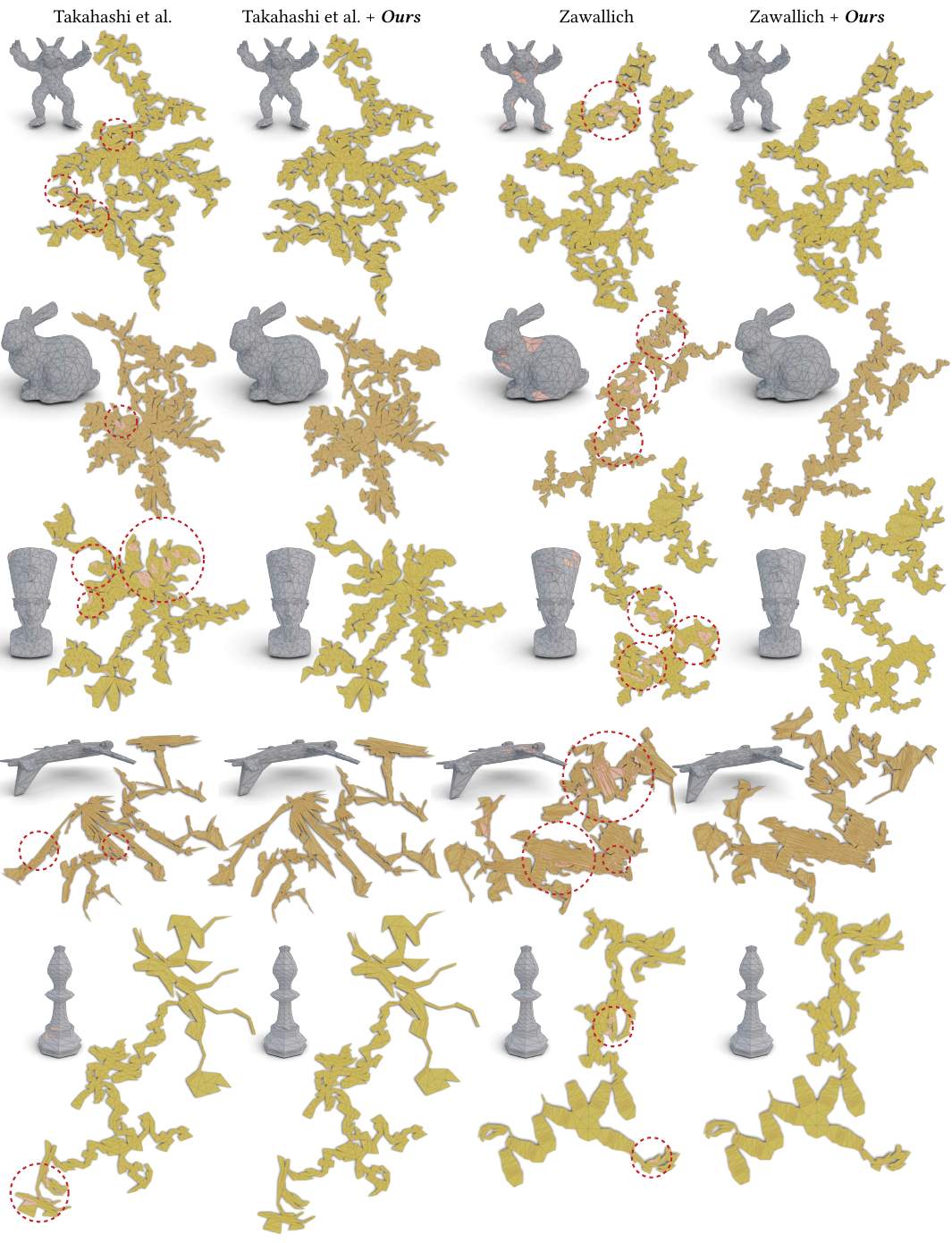}
\vspace{-0.3cm}
    \caption{For all relatively simple meshes in this gallery, the methods by \cite{takahashi2011optimized} and \cite{zawallich2024unfolding} fail to find a single-patch, non-overlapping unfolding. Our algorithm minimally modifies the mesh to accomplish this in every case. \textbf{See accompanying video} and quantitative results in \autoref{table:statistics-gallery}} 
\label{fig:gallery-example}
\end{figure*}

Our algorithm is composed of several sequential steps. As we show in \autoref{fig:ablation}, in which we highlight overlapping triangles and use Chamfer distance as a proxy for the average approximation error, the combination of all of these steps balances the lack of overlap with the quality of the output's approximation.
Like the unfoldable tetrahedron in \autoref{fig:demaine}, \autoref{fig:ablation} is a didactic example. However, we also qualitatively showcase the proficiency of our method over several complex examples in \autoref{fig:teaser} (using \cite{takahashi2011optimized} as initializer) and \autoref{fig:gallery-example} (with \textbf{accompanying video}), for which quantitative results are also reported in \autoref{table:statistics-gallery}.

In particular, there are two algorithmic choices that we choose to evaluate in more detail: namely, the vertex manipulation in \autoref{section:vertex-repositioning} and the heuristic placement of the collapsed vertex in \autoref{section:mesh-simplification} (see \autoref{table:statistics-variants} in Supplemental), where we report the metrics of different variants of our method for our two datasets). As expected, we find that the vertex manipulation step consistently leads to better results, while a more careful choice of the collapsed vertex generally leads to a lower approximation error.

\vspace{-0.1cm}
\subsection{Comparisons}\label{section:results-comparisons}

By relaxing the problem and allowing for minimal changes in the input geometry, we dramatically increase the success rate of previous works, as we show quantitatively in \autoref{table:many-shapes-experiment} and qualitatively in \autoref{fig:related-fig} (the bar chart corresponding to our \emph{fine} dataset).

As shown in \autoref{table:many-shapes-experiment}, on our \emph{coarse} dataset, the algorithm by \cite{takahashi2011optimized} successfully unfolds 40.24\% of the meshes; when using their attempts as initializers for our method, we more than double that success rate into 92.42\%. The effect is even more pronounced on the \emph{fine} dataset, where the success rate increases eight-fold, from 10.55\% to 84.10\%.
Our algorithm even outperforms the concurrent work by \cite{zawallich2024unfolding} where they use their unfolder \cite{Zawallich2023}, which achieves a success rate of 85.30\% on the \emph{coarse} dataset (compared to our 91.96\%) and 62.99\% on the \emph{fine} one (compared to our 87.93\%). 

Our unfolding-aware decimation strategy is crucial. To validate this, we performed a further test on the subset of meshes which are unfolded by our method but not by any of the related works. If the meshes are decimated using standard techniques (e.g., \textit{qslim} \cite{QEMDecimation1997}) to the same refinement level, they are most often still non-unfoldable. For roughly 45\% of meshes in our \emph{coarse} subset dataset and 55\% of the \emph{fine} subset dataset both \cite{takahashi2011optimized} and \cite{Zawallich2023} failed.

\subsection{Applications \& Generalizations}

Our method can be readily applied to common manufacturing frameworks, as we showcase with our fabricated results in \autoref{fig:fabrication-results}. For this figure, initially unfoldable input meshes of a fox and an Italian plumber are simplified by our algorithm and then fabricated from paper, with the aid of a 3D printed guide.

\begin{figure}
\includegraphics{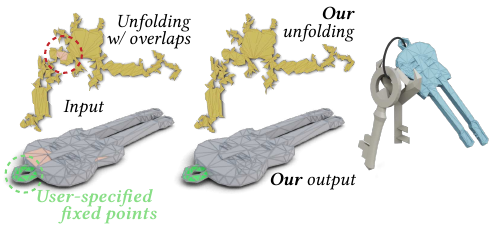}
\vspace{-0.35cm}
\centering
\caption{Fixed-point constraints can be easily incorporated into our algorithm; for example, to ensure that this guitar-shaped keychain is functional.}
\label{fig:guitar}
\vspace{-0.35cm}
\end{figure}

The mesh simplification carried out by our method may be undesirable in some fabrication settings: for example, when one wishes to preserve certain features because of their functionality. Our algorithm can be easily modified to incorporate these fixed points (by simply not allowing them to be collapsed or displaced), as we showcase by simulating a fabricated keychain in \autoref{fig:guitar}.

\vspace{-0.1cm}
\section{Conclusions \& Future Work}\label{section:conclusions}

We have introduced a computational approach for finding an approximate mesh of a given 3D shape that admits a single-patch, non-overlapping unfolding.
We propose doing this via a custom mesh processing algorithm that combines unfolding-aware vertex manipulation and mesh decimation to iteratively modify a mesh with an overlapping unfolding into a mesh with a non-overlapping one.
As we have shown, our approach can be used to dramatically increase the range of shapes that can be succesfully unfolded in a single patch, making it particularly promising for applications in fabrication and design.

Despite its improvement on previous approaches, our method still occasionally fails to find a valid unfolding for certain inputs, instead getting `stuck' at a stage without valid edge collapses or vertex displacements. 
In these cases, one may consider discarding the current spanning tree and starting over with a newly computed one. In \autoref{table:statistics-reinitialisation} (Supplemental), we experiment with a reinitialization strategy that does just that, and show that it can improve the success rate of our algorithm. 
Nonetheless, this comes at a significant cost, as re-computing a high-quality spanning tree is computationally expensive.
Instead, future work may consider less agressive, more efficient reinitialization strategies, or even ways of avoiding getting stuck in the first place.

The most promising avenues for future work relate to making our method even more directly applicable to fabrication settings.
For example: our algorithm is aimed at producing single-patch unfoldings. However, for large numbers of triangles, this can be impractical, and future work may consider allowing a user to specify a particular number of patches.
Similarly, fabricating objects with paper can lead to gluing a lot of edges even if it is single component. To ease in this task, one can integrate the glue tabs proposed by \cite{takahashi2011optimized} and \cite{korpitsch2020simulated} on our output.

From a more general perspective, we hope that the lessons in this work can serve to develop new unfolding algorithms that are combined with mesh processing techniques: for example, by ensuring that overlaps occur in regions of smooth curvature where local decimation would not affect the overall mesh quality too much, or in convex regions on which overlaps are easier to resolve.
Similarly, we believe that more sophisticated edge collapse strategies (for example, those that consider the consequences of several sequential collapses at a time instead of a single one) may improve the robustness of unfolding-aware decimation algorithms like ours.

Finally, our algorithm sits in a more general class of methods that \emph{relax} known-to-be-hard problems into easier, approximate ones that are equally valid for the needs of a given application.
We believe fabrication and computational design are particularly promising application realms for this kind of approach, and hope our work can inspire others to explore similar strategies.

\bibliographystyle{eg-alpha-doi}
\bibliography{references}

\newpage

\newpage
\clearpage

\appendix

\section{Pseudocode}

This appendix provides pseudocode for the main steps of our algorithm, to aid in their implementation. \autoref{algorithm:algorithm-main} shows the main outer loop of our method, while \autoref{algorithm:decimation} covers the unfolding-aware decimation strategy in \autoref{section:mesh-simplification} step by step.

\begin{algorithm}
\vspace{0.2cm}
    \caption{Main Algorithm}\label{algorithm:algorithm-main}
    \begin{algorithmic}[1]
    \State Input: Manifold mesh ($\Omega^0$)
    \State Output: Manifold mesh ($\Omega^\star$) and a simple unfolding ($\Gamma^\star$)
    \State Generate the initial unfolding ($\Gamma^0$) (Section~\ref{section:initialization})
    \State $iter \gets 0$
    \State numOverlaps = computeOverlaps($\Gamma^0$)
    \While {numOverlaps > 0 or AND iter $<$ maxIter}
        \State Unfolding-aware vertex manipulation. (Section~\ref{section:vertex-repositioning})
        \State Unfolding-aware mesh decimation. (Section~\ref{section:mesh-simplification})    
        \State $iter \gets iter + 1$ (Section~\ref{section:main-method})
        \State numOverlaps = computeOverlaps($\Gamma^i$)
    \EndWhile
    \State Post-process the mesh. (Section~\ref{section:postprocessing})
    \end{algorithmic}
    \end{algorithm}

    \begin{algorithm}
    \vspace{0.2cm}
        \caption{Spanning tree-preserving edge collapse}\label{algorithm:decimation}
        \begin{algorithmic}[1]
            \State Generate the priority queue for edges to collapse
            \While {priority queue != empty}
                \State edge $\gets$ priorityQueue.pop() 
                \State collapse(edge) as explained in section \ref{section:mesh-simplification}
                \If{edge collapse generate degenerate features} 
                    \State Revert
                \EndIf    
                \State compute unfolding from the spanning tree 
                \State count overlapping triangles in the unfolding
                \If{overlap count not decreased.} 
                    \State Revert
                \Else
                    \State confirm edge collapse.
                \EndIf
                \If{(priority queue == empty)}
                    \If{exit-strategy == \textit{broadening-our-search} and \textit{iter} $>$ 1}
                        \State{Add 1-vertex ring edges of overlapping edges as well.}
                    \EndIf
                    \If{exit-strategy == \textit{lowering-our-expectations}}
                        \State{remove the Edge that causes least new overlaps in 2D.}
                    \EndIf
                \EndIf
        \EndWhile
        \State end
            
    \end{algorithmic}
\end{algorithm}

\begin{figure}[b]
    \includegraphics{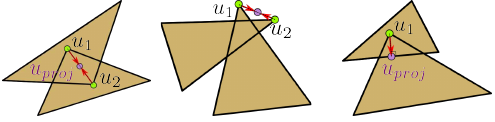}
    \caption{We showcase the cases of overlapping triangles that we attempt to resolve by our vertex repositioning approach. The figure contains the target projected vertex that could potentially resolve the overlap.}
    \label{fig:flow-didactic}
    \vspace{0.3cm}
\end{figure}

\section{Surface flow for resolving overlaps}
\label{section:surface-flow}
    
For each pair of overlapping triangles, we compute the projected vertex $u_{proj}$ (see Figure~\ref{fig:flow-didactic}). The projection points $u_{proj}$ are then lifted in 3D using barycentric coordinates to obtain $v_{proj}$. We define
$$\Phi_c = \frac{1}{2}\sum_{k=1}^n ||v_k - v_{k, proj}||^2\text{.}$$ Furthermore, to ensure that triangles do not flip or drastically change their shape, we add regularising energy $\Phi_s$ that helps maintain the shape of the triangles: $$\Phi_s = \frac{1}{2}\sum_{h=1}^m A_h E_h:E_h$$ where $A_h$ is the area of triangle $h$ from $\Gamma^t$ and $E_h$ is the Green strain tensor of the  deformation between the unfolding $\Gamma^t$ and the current 3D mesh. The Green strain $E$ is discretized by assuming a constant strain over each triangle. More details on the computation and differentiation of $E$ can be found in the course by \cite{femdefo}. We perform a gradient descent to optimize the energy $\Phi = w_c\Phi_c+w_s\Phi_s$ over the vertices $v_k$ of the 3D mesh. At each iteration of the surface flow, we displace each vertex $v_k$ of $d_k = \delta t \frac{\partial\Phi(t)}{\partial v_k}$. We use 100 iterations, and set $w_c = 1$ and $w_s = 10$ with a step $\delta t = 0.01$ for all our experiments.

\section{Post-processing details}
\label{section:post-processing-appendix}

Our post-processing energy is composed of three terms. The \emph{distance} term takes the form
\begin{equation}
    E_d(\Omega,\Omega^0) = \frac{1}{2} \sum_{v_i \in \Omega} \|v_i - \projection(v_i,\Omega^0)\|^2 + \sum_{w_j \in \Omega^0} \|w_j - \projection(w_j,\Omega)\|^2\text{,}
\end{equation}
which is linearized by making $p_i =  \projection(v_i,\Omega^0)$ (ignoring the dependency with $v_i$) and $\projection(w_j,\Omega_t) = \mathbf{B}_j\mathbf{w}$, where $\mathbf{B}_j$ contains the barycentric coordinates of the projection on $\Omega_t$ and $\mathbf{w}$ is a matrix vertically concatenating the coordinates of every vertex in $\Omega_t$.
The \emph{overlap} term is defined as
\begin{equation}
    E_o(\Gamma) = E_1(\Gamma) + E_2(\Gamma)\text{,}
\end{equation}
where $E_1$ is the 2D collision energy
\begin{equation}
    E_1(\Gamma) = \frac{1}{2}\sum_{i=1}^n ||u_i - u_{i, proj}||^2
\end{equation}
and $E_2$ is a 2D regularizer
\begin{equation}
    E_2(\Gamma) = \frac{1}{2}\sum_{j=1}^m A_j E^{2d}_j:E^{2d}_j\text{,}
\end{equation}
where the strain tensor $E^{2d}$ is computed for the deformation between the reference state $\Gamma$ and the current state during the gradient descent.
Finally, the \emph{coupling} term is given by
\begin{equation}
    E_c(\Omega,\Gamma) = \frac{1}{2}\sum_{j=1}^m A_j E^{2d3d}_j:E^{2d3d}_j\text{,}
\end{equation}
where $E^{2d3d}$ is the Green strain of the deformation from $\Gamma$ to $\Omega$.
These terms are combined into the post-processing energy $E_p(\Omega,\Gamma)$ as described in \autoref{section:postprocessing}.

\begin{table*}
    \begin{center}
    \caption{Our algorithm is most successful when incorporating the vertex repositioning strategy described in \autoref{section:vertex-repositioning}. On the other hand, the impact of the edge collapse heuristic described in \autoref{section:mesh-simplification} depends on the chosen initialization strategy. In all cases, our post-processing significantly improves mesh quality.} \label{table:statistics-variants}
    \resizebox{\linewidth}{!} 
    {
    \addtolength{\tabcolsep}{2.5pt}
    \renewcommand{\arraystretch}{1.1} %
    \begin{tabular}{l||c|c|c||c|c|c|c|c}
    \toprule
    \rowcolor{tableheader}
    Dataset & Initialization & \makecell{`Smarter'\\edge collapse} & \makecell{Vertex\\Repositioning} & Success & Runtime & \makecell{Chmf error} & \makecell{Post-process\\runtime} & \makecell{Post-process\\improvement} \\
    \midrule
    \emph{coarse} & Takahashi & No & No & 92.06\% & 1.11 s & 0.0017 & 0.37 s & 6.91\% \\
    \rowcolor[HTML]{EFEFEF} \emph{coarse} & Takahashi & No & Yes  & 91.64\% & 3.14 s & 0.0017 & 0.39 s & 7.15\% \\
    \emph{coarse} & Takahashi & Yes & No  & 91.70\% & 3.89 s & 0.0015 & 0.36 s & 7.05\% \\
    \rowcolor[HTML]{EFEFEF} \emph{coarse} & Takahashi & \textbf{Yes} & \textbf{Yes}  & \textbf{92.42\%} & \textbf{2.77 s} & \textbf{0.0015} & \textbf{0.36 s} & \textbf{4.11\%} \\
    \midrule
    \emph{fine} & Takahashi & No & No & 83.11\% & 13.80 s & 0.0018 & 0.93 s & 5.11\% \\
    \rowcolor[HTML]{EFEFEF} \emph{fine} & Takahashi & No & Yes  & 83.57\% & 25.35 s & 0.0018 & 0.92 s & 5.03\% \\
    \emph{fine} & Takahashi & Yes & No  & 83.89\% & 12.96 s & 0.0018 & 0.93 s & 2.36\% \\
    \rowcolor[HTML]{EFEFEF} \emph{fine} & Takahashi & \textbf{Yes} & \textbf{Yes}  & \textbf{84.10\%} & \textbf{18.64 s} & \textbf{0.0015} & \textbf{0.76 s} & \textbf{3.31\%} \\
    \midrule
    \emph{coarse} & Zawalich & No & No & 91.87\% & 5.33 s & 0.0050 & 0.30 s & 1.57\% \\
    \rowcolor[HTML]{EFEFEF} \emph{coarse} & Zawalich & \textbf{No} & \textbf{Yes}  & \textbf{91.96\%} & \textbf{11.69 s} & \textbf{0.0050} & \textbf{0.28 s} & \textbf{3.26\%} \\
    \emph{coarse} & Zawalich & Yes & No  & 91.18\% & 5.39 s & 0.0048 & 0.29 s & 3.26\% \\
    \rowcolor[HTML]{EFEFEF} \emph{coarse} & Zawalich & Yes & Yes  & 91.21\% & 8.15 s & 0.0045 & 0.29 s & 1.73\% \\
    \midrule
    \emph{fine} & Zawalich & No & No & 87.56\% & 7.52 s & 0.0012 & 0.77 s & 7.26\% \\
    \rowcolor[HTML]{EFEFEF} \emph{fine} & Zawalich & \textbf{No} & \textbf{Yes}  & \textbf{87.93\%} & \textbf{14.52 s} & \textbf{0.0011} & \textbf{0.76 s} & \textbf{0.87\%} \\
    \emph{fine} & Zawalich & Yes & No  & 86.95\% & 5.85 s & 0.0011 & 0.76 s & 1.76\% \\
    \rowcolor[HTML]{EFEFEF} \emph{fine} & Zawalich & Yes & Yes  & 87.28\% & 9.47 s & 0.0011 & 0.77 s & 1.92\% \\
    \midrule
    \end{tabular}
    }
    \end{center}
\end{table*}

\begin{table*}
    \begin{center}
    \caption{Occassionally, our algorithm fails to find a solution. In such cases, we discard the current spanning tree and instead re-run our entire pipeline, treating the last iteration's mesh as the new input. This strategy results in a higher success rate, but at the cost of increased runtime.}\label{table:statistics-reinitialisation}
    \resizebox{\linewidth}{!} 
    {
    \addtolength{\tabcolsep}{5pt}
    \renewcommand{\arraystretch}{1.1} %
    \begin{tabular}{l||c||c|c||c|c}
        \toprule
    \rowcolor{tableheader}
    Dataset & Initialization & \makecell{Success\\(without re-initialization)} & \makecell{Success\\(with re-initialization)} & \makecell{Time spent\\in our algorithm} & \makecell{Time spent\\in initializers} \\
    \midrule
    \emph{coarse} & Takahashi & 92.42\% & \textbf{99.02\%} & 135.65 s & 82.29 s \\
    \rowcolor[HTML]{EFEFEF} \emph{coarse} & Zawalich & 91.96\% & \textbf{97.44\%} & 141.00 s & 15.82 s \\
    \emph{fine} & Takahashi & 84.10\% &\textbf{95.33\%} & 660.68 s & 367.58 s \\
    \rowcolor[HTML]{EFEFEF} \emph{fine} & Zawalich & 87.93\% & \textbf{91.39\%} & 423.43s & 114.34 s \\
    \midrule
    \end{tabular}
    }
    \end{center}
\end{table*}

\section{Additional Results}

We extensively evaluated our method on our two datasets with experiments beyond those contained in the main paper, which we provide in this section.
\autoref{table:statistics-variants} shows the effects of two specific algorithmic choices: namely, the use of the vertex repositioning strategy and the edge collapse heuristic (see discussion in \autoref{section:results-comparisons}). 
\autoref{table:statistics-reinitialisation} demonstrates the impact of re-initializing our algorithm when it fails to find a solution (see discussion in \autoref{section:conclusions}).

\end{document}